\documentclass[final,5p,times,twocolumn]{elsarticle}

\usepackage{natbib}
\usepackage[flushleft]{threeparttable}
\usepackage{graphicx}
\usepackage[english]{babel}
\usepackage{color}
\usepackage{xcolor}
\usepackage{wrapfig}
\graphicspath{{images/}}
\usepackage{csquotes}
\usepackage[version=3]{mhchem} 
\usepackage{graphicx,amstext,amsmath}
\usepackage{mhchem}
\usepackage{array}
\usepackage{mathrsfs}
\usepackage{comment}
\usepackage{ulem}

\usepackage{xr}
\makeatletter
\newcommand*{\addFileDependency}[1]{
  \typeout{(#1)}
  \@addtofilelist{#1}
  \IfFileExists{#1}{}{\typeout{No file #1.}}
}
\makeatother

\newcommand*{\myexternaldocument}[1]{
    \externaldocument{#1}
    \addFileDependency{#1.tex}
    \addFileDependency{#1.aux}
}

\myexternaldocument{SI}
\usepackage{graphicx}
\usepackage{floatrow}
\usepackage{subfig}

\usepackage{amssymb}
\usepackage{amsthm}

\biboptions{sort&compress,super}

\journal{JACS}

\begin{document}
\begin{frontmatter}
\title{P1 center electron spin clusters are prevalent in type Ib diamond}

\author{Santiago Bussandri$^{1}$}
\author{Daphna Shimon$^{2}$}
\author{Asif Equbal$^{3,4}$}
\author{Yuhang Ren$^{5}$}
\author{Susumu Takahashi$^{5,6}$}
\author{Chandrasekhar Ramanathan$^7$}
\author{Songi Han$^{1,8,9,*}$}
\ead{songi@chem.ucsb.edu}
\address{$^{1}$Department of Chemistry and Biochemistry, University of California, Santa Barbara, Santa Barbara, California 93106, United States.}
\address{$^{2}$Institute of Chemistry, The Hebrew University of Jerusalem, Edmond J. Safra, Givat Ram, Jerusalem 9190401, Israel.}
\address{$^{3}$Department of Chemistry, New York University, Abu Dhabi 129188, United Arab Emirates.}
\address{$^{4}$Center for Quantum and Topological Systems, New York University, Abu Dhabi 129188, United Arab Emirates.}
\address{$^{5}$Department of Physics and Astronomy, University of Southern California, Los Angeles, California 90089, USA.}
\address{$^{6}$Department of Chemistry, University of Southern California, Los Angeles, California 90089, USA.}
\address{$^{7}$Department of Physics and Astronomy, Dartmouth College, Hanover, New Hampshire 03755, USA.}
\address{$^{8}$Department of Chemical Engineering, University of California, Santa Barbara, California 93106, USA.}
\address{$^{9}$Department of Chemistry, Northwestern University, Evanston, Illinois 600208, USA.}

\begin{abstract}
Understanding the spatial distribution of P1 centers is crucial for diamond-based sensors and quantum devices. P1 centers serve as a polarization source for DNP quantum sensing and play a significant role in the relaxation of NV centers. Additionally, the distribution of NV centers correlates with the distribution of P1 centers, as NV centers are formed through the conversion of P1 centers. We utilized dynamic nuclear polarization (DNP) and pulsed electron paramagnetic resonance (EPR) techniques that revealed strong clustering of a significant population of P1 centers that exhibit exchange coupling and produce asymmetric lineshapes. The $^{13}$C DNP frequency profile at high magnetic field revealed a pattern that requires an asymmetric EPR lineshape of the P1 clusters with electron-electron (e-e) coupling strengths exceeding the $^{13}$C nuclear Larmor frequency. EPR and DNP characterization at high magnetic fields was necessary to resolve energy contributions from different e-e couplings. We employed a two-frequency pump-probe pulsed Electron Double Resonance (ELDOR) technique to show crosstalk between the isolated and clustered P1 centers. This finding implies that the clustered P1 centers affect all P1 populations. Direct observation of clustered P1 centers and their asymmetric lineshape is a novel and crucial insight into understanding magnetic noise sources for quantum information applications of diamonds and for designing diamond-based polarizing agents with optimized DNP efficiency for $^{13}$C and other nuclear spins of analytes. We propose that room temperature $^{13}$C DNP at high field, achievable through straightforward modifications to existing solution-state NMR systems, is a potent tool for evaluating and controlling diamond defects.
\end{abstract}


\end{frontmatter}


\section{Introduction} \label{Introduction}

Defects in diamonds have become increasingly important in a variety of fields, from quantum computing to quantum sensing. 
One of the most widely studied defects is the nitrogen vacancy (NV) center, which consists of a substitutional nitrogen atom and an adjacent vacancy in the diamond lattice~\cite{Gruber1997, Degen2008, Maze2008, Balasubramanian2008, Taylor2008}. 
The NV center has been found to exhibit remarkable properties, including long spin coherence times, room temperature operation, and optical addressability~\cite{Jelezko2004, Epstein2005, Childress2006, Takahashi2008, Balasubramanian2009}. As a result, it has been used in a range of applications, such as single-photon sources, nanoscale magnetic sensors, and quantum memories~\cite{10.1063/5.0138050, RevModPhys.92.015004, Laraoui2013, Mamin2013, Staudacher2013, DeLange2012, Grinolds2014, Shi2015, Abeywardana2016, 10.1063/5.0006014, PhysRevX.9.031045}. 
NV centers arise from P1 centers that are single substitutional nitrogen defects in the diamond lattice that also have a long coherence time, even at room temperature~\cite{loubser_electron_1978, Takahashi2008}. 
Studying the spatial distribution of P1 and NV centers is critical for many reasons. For example, P1 centers are a major source of polarization for nearby carbon nuclei that can be used for quantum sensing applications~\cite{doi:10.1126/sciadv.aar5492, doi:10.1073/pnas.1807125115, doi:10.1073/pnas.1908780116, shimon_large_2022}. 
P1 centers are also the main culprit for the relaxation of the NV center EPR signal and coherence, and hence may be the bottleneck in applications where long coherence times from the NV center are needed for sensing applications~\cite{vanWyk97, Takahashi2008, stepanov_determination_2016, bauch2020decoherence, li_determination_2021}. 
Knowledge of the microscopic spin distribution is essential for the investigation of exotic physics in disordered dipolar spin systems \cite{choi_observation_2017, Kucsko18}. {Diamond is also a promising polarization agent for dynamic nuclear polarization (DNP) for achieving high polarization of intrinsic $^{13}$C nuclear spins for sensing or imaging applications~\cite{Ajoy2018, Meriles2019, Boele2020, Pagliero2020} or of nuclear spins of external analytes~\cite{Rej2017}.}

The formation of NV centers is often realized by the application of high-energy electron radiation to create vacancies and subsequent high-temperature annealing processes where the mobile vacancies migrate through the lattice until they meet with the nitrogen defects to form the NV centers. 
Therefore, the distribution of the NV centers is closely related to the original distribution of the P1 centers.

While nitrogen defects can exist in both natural and synthetic diamond crystals, synthetic diamonds have emerged as a key tool for scientific research, due to the tunability of their properties and the ability to generate controlled impurities. High-pressure high-temperature (HPHT) and chemical vapor deposition (CVD) are the two main methods used to produce synthetic diamonds. The vast majority of synthetic diamond is made using the HPHT method. This procedure aims to mimic the thermodynamic conditions in which diamonds form naturally. The nitrogen is incorporated into the diamond lattice from the atmosphere and growth materials.
For this reason, HPHT diamonds usually present a higher percentage of impurities with less controlled spatial distribution. CVD diamonds, on the other hand, are grown by exposing a substrate to hydrocarbon and hydrogen gas in a vacuum chamber. Nitrogen impurities can be introduced into the diamond lattice by adding a controlled amount of nitrogen gas to the chamber. CVD diamonds are often preferred for certain applications that require single-spin localization due to their high purity, low nitrogen concentration, and potentially even the direct formation of NV centers~\cite{Ishiwata17}. CVD diamonds, with their low nitrogen concentration (generally lower than 5 ppm) received the name of type IIa diamonds. However, for ensemble measurements, diamonds with a concentration higher than 10 ppm are used and classified as type Ib diamonds, usually made by HPHT procedures. 
In case of both type Ib and type IIa diamonds, the distribution of P1 centers in the diamond and the conversion efficiency to NV centers are vital to their performance for quantum sensing and spectroscopy. Even if the conversion rate from P1 to the NV center is low enough to obtain isolated NV centers, the remaining P1 centers will be a main source of relaxation for the NV centers. For this reason, it is critical to develop methods to characterize interactions or clustering of P1 and NV centers. In our study, we focus on characterizing representative type Ib diamonds to test high-field DNP- and EPR-based methods to evaluate P1-P1 interactions. In fact, this study was inspired by a surprising observation reported in the literature that hinted towards the existence of nitrogen defect clusters in HPHT-made, type Ib diamonds~\cite{shimon_large_2022}. 

Electron paramagnetic resonance (EPR) spectroscopy has been a powerful tool to identify paramagnetic impurities in diamonds~\cite{loubser_electron_1978, smith_electron-spin_1959}.
Loubser and van Wyk summarized the EPR identification of more than 40 paramagnetic impurities in diamonds, consisting of a single and a few defects~\cite{loubser_electron_1978}.
The spin concentration of paramagnetic impurities can be studied by the analysis of the EPR intensity/linewidth~\cite{Samsonenko1965, smith_distribution_1966, loubser_electron_1978}, spin relaxation times ($T_2$ and $T_1$)~\cite{PhysRev.74.1168, Abragam, Samsonenko1965, smith_distribution_1966, loubser_electron_1978, Wang2013, bauch2020decoherence} and the electron-electron double resonance (ELDOR) spectroscopy~\cite{stepanov_determination_2016,li_determination_2021}.
However, when the sample has inhomogeneous spatial distributions of the paramagnetic centers and multiple spin species with a wide range of the spin concentrations in the detection volume, the determination is challenging because these analyses are based on the knowledge of the spin properties, including the type of spins, spectral distributions and a type of couplings to surrounding spins, {\it e.g.} dipolar or exchange (J) coupling.
For example, Li {\it et al.}, has studied the local spin concentrations of a diamond crystal with highly inhomogeneous spin concentration and the relationship between $T_2$ and the P1 concentration using a combination of NV-detected ELDOR and $T_2$-based methods, and showed that some diamond domains exhibit short $T_2$ values which cannot be explained by the nominal P1 concentration~\cite{li_determination_2021}. Still, the origin of the additional contribution was not identified.
Shenderova {\it et al.} discussed, in a review article, the occurrence and potential application of different color centers in diamonds generated by migration and strong coupling between two nitrogen defects and NV centers that give rise to EPR signatures representing forbidden transitions at half field. However, the relationship between clustered P1 and the various color centers, and whether the majority P1 populations communicate with these color centers is unknown~\cite{Shenderova2019}.
We turn to DNP and ELDOR spectroscopy to study the cross talk between different P1 populations via the $^{13}$C NMR probe frequency or directly via pump probe experiments, respectively. DNP is a widely used technique to enhance the intensity of the nuclear magnetic resonance (NMR) signal, but our interest in DNP stems from its sensitivity to the nature of interacting spins~\cite{ equbal_truncated_2018, shimon_large_2022}.
A recent DNP study of type-Ib diamond samples by Shimon discussed that the observed truncated cross effect can be explained by the existence of coupled P1 spins in the sample in addition to isolated P1 center~\cite{shimon_large_2022}, { but the fundamental properties of the coupling, whether they are driven by dipolar or exchange coupling, whether there is cross talk between the different P1 population, the population of clustered P1 centers and the mechanistic origin of the proposed Overhauser DNP effect remained unknown.} 

In this study, we uncover the co-existence of clustered and isolated P1 centers in type-Ib diamonds using a combination of multi-frequency DNP, pulsed high-frequency EPR, and ELDOR techniques. We start with the observation of the frequency dependant DNP profile of carbon-13 ($^{13}$C) enhancement which turns out to be a very sensitive probe of electron spin interactions and clustering. With insight from DNP at hand, we perform EPR lineshape analysis that reveals the existence of two components of isolated and clustered species, as well as EPR nutation frequency analysis that reveals the high-spin nature of the clustered P1 species. Finally, we interrogate the crosstalk between these species by pump-probe ELDOR experiments and discuss how these findings are broadly relevant to the commercially available type Ib diamonds, widely used for ensemble sensing and DNP applications.

\section{Results and Discussion}

\subsection{ $^{13}$C DNP Experiments}

\begin{figure*}[h]
    \centering
    \includegraphics[width =\textwidth]{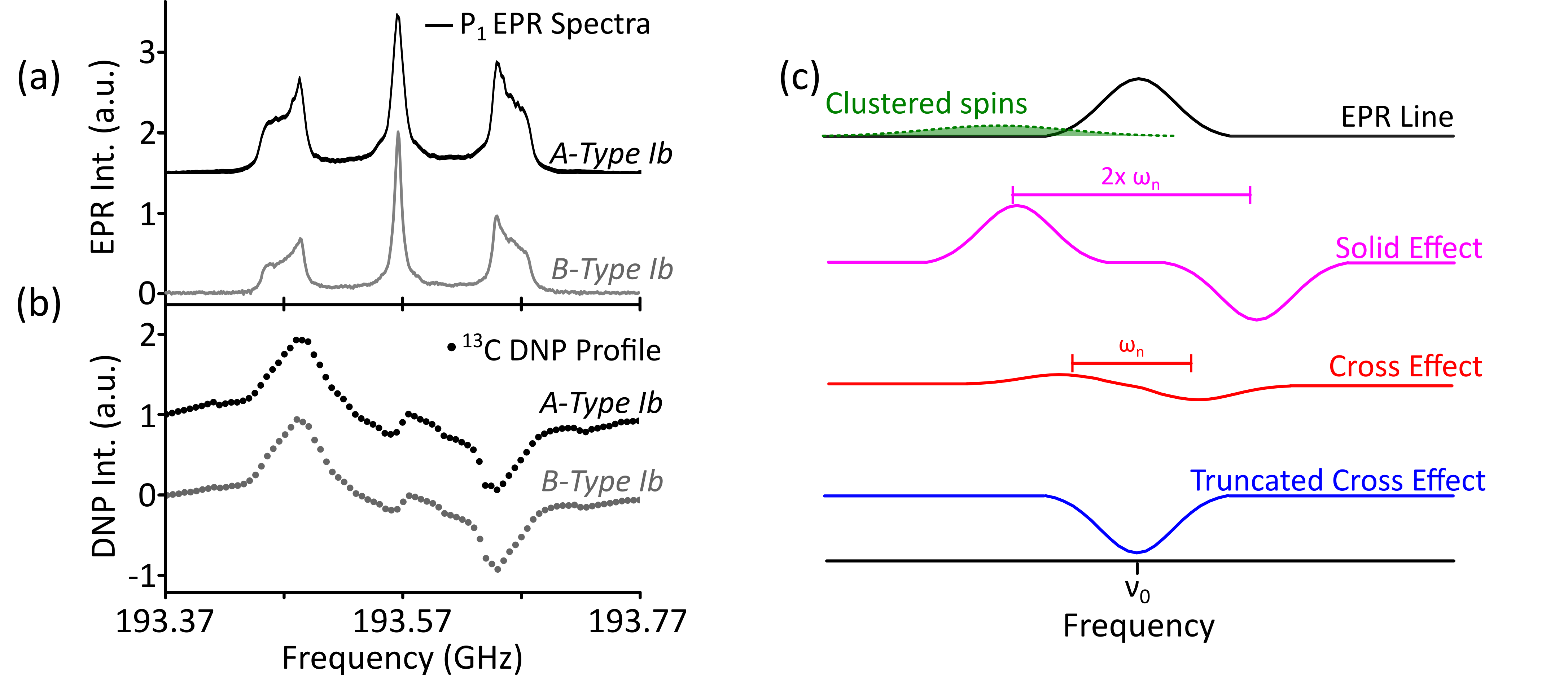}
    \caption{a) P1 echo-detected frequency stepped spectrum measured from two HPHT, type Ib microdiamond samples (A and B). b) $^{13}$C DNP profile of Sample A and B. The data in a) and b) were measured at room temperature, 7 T, and static conditions. c) Schematic of possible DNP mechanisms that contribute to the DNP profile. The hypothetical EPR line is shown in black with a line broadening of the order of the nuclear Larmor frequency ($\omega_n$). Solid effect (shown in pink) develops two antiphase peaks with the same lineshape as the EPR line separated by 2 times $\omega_n$. Cross effect (shown in red) presents two antiphase peaks separated by $\omega_n$. The truncated cross effect (shown in blue) arises from the interaction of the hypothetical EPR feature with clustered spins that present a broad and shifted spectrum (shown in green). It results in a unique peak with a negative sign due to the relative position of the broadened EPR line to the original EPR spectrum.}
    \label{DNP Samples}
\end{figure*}

This study focuses on detecting clustered P1 centers in commercially available and widely used type Ib diamonds. Figure \ref{DNP Samples}a exhibits the echo-detected frequency stepped spectrum of the P1 center measured from two HPHT, type Ib microdiamond powder under static (non-spinning) conditions at 7 T and room temperature. The observed spectres feature a central peak with an FWHM of 11 MHz and two side peaks, with of FWHM of 32 MHz, with a separation of 82 MHz between them. A hyperfine coupling of the P1 center with the $I=1$ Nitrogen nucleus is responsible for this observed three-peak manifold. The asymmetry in the intensities of the $m_s=\pm 1$ hyperfine peaks is due to a small g-anisotropy which will be quantified with corresponding experiments at high magnetic fields.

\begin{figure*}[h]
    \centering
    \includegraphics[width =0.95\textwidth]{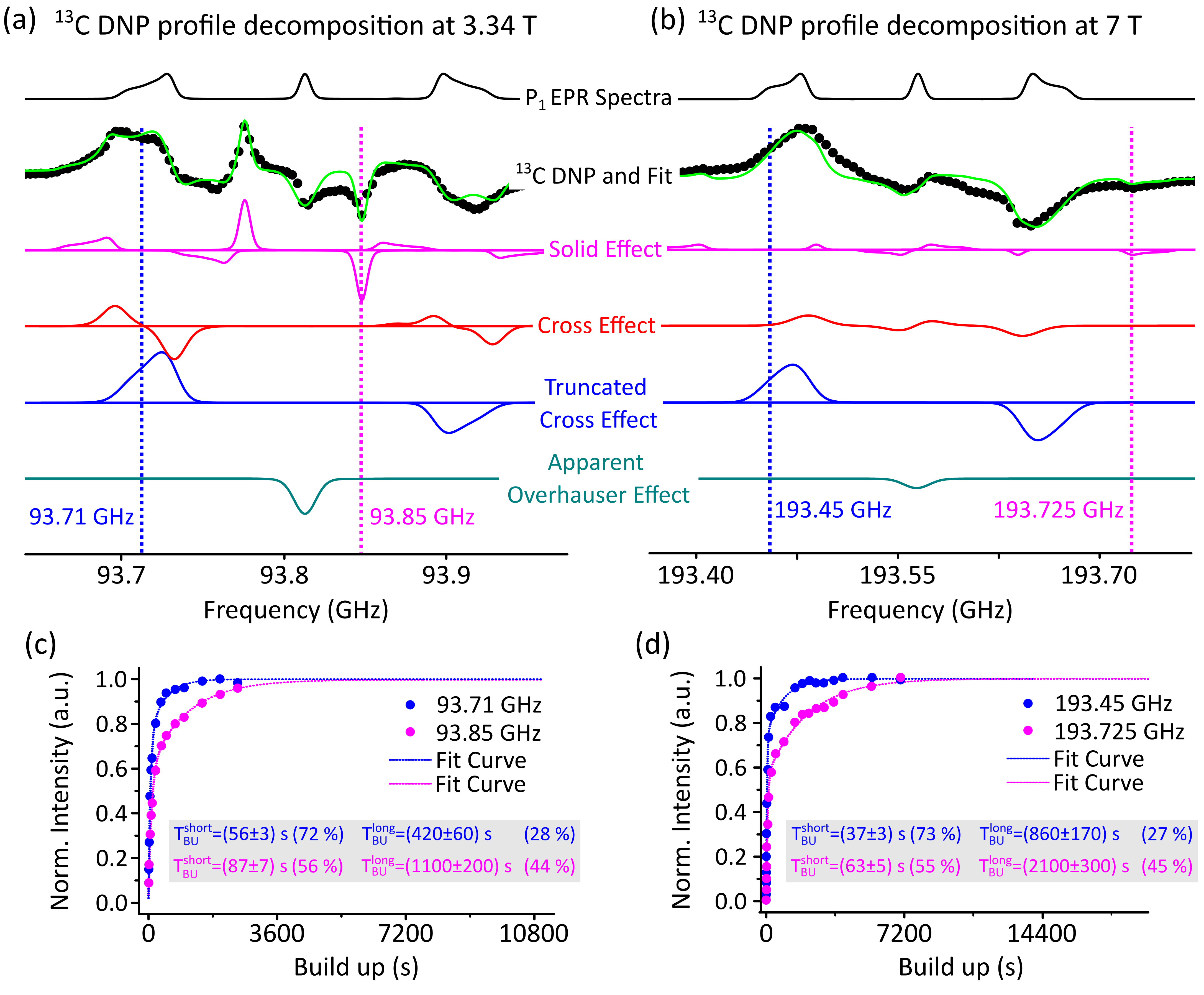}
    \caption{P1 EPR spectra, $^{13}$C DNP profile, and fit, and decomposition into four different DNP mechanisms at room temperature and at two different magnetic fields, a) 3.34 T and b) 7 T. For the $^{13}$C DNP profile, 3000 s of MW irradiation was used at both magnetic fields. Blue/magenta dashed lines highlight the frequencies at which the tCE/SE feature is dominant. c) $^{13}$C DNP build-up time and fittings (dotted line) measured at 3.34 T at two different frequencies (93.71 GHz and 93.85 GHz). d) $^{13}$C DNP build-up time and fittings (dotted line) measured at 7 T at 193.45 GHz and 193.725 GHz. Figure a is adapted from "Large Room Temperature Bulk DNP of 13C via P1 Centers in Diamond" D. Shimon, K. A. Cantwell, L. Joseph, E. Q. Williams, Z. Peng, S. Takahashi, and C. Ramanathan. The Journal of Physical Chemistry C 2022 126 (41), 17777-17787. COPYRIGHT 2022 American Chemical Society.}
    \label{13C_DNP_DECOMPOSITION}
\end{figure*}

The existence of clustered spins is generally difficult to directly identify through echo-detected EPR measurements because they tend to be buried within the EPR lines and have faster relaxation times. { Critically, even if they are visible as broadening or an elevated baseline in an EPR spectrum, contribution from specific clusters versus other paramagnetic impurities cannot be easily differentiated.}
Interestingly, DNP frequency profiles reveal richer features than EPR spectra, and can more readily identify clustered P1 centers { because the mechanisms relying on an isolated electron spin giving rise to the solid effect (SE) DNP versus two or more coupled electron spins giving rise to cross effect (CE) type DNP generate characteristic and identifying features.} Although high-field DNP experiments have not been commonly used for the study of nitrogen defect centers, a recent study~\cite{shimon_large_2022} showed that the extra features of the experimental $^{13}$C DNP frequency profile, that appear on top of the signatures of DNP effects originating from isolated P1 centers, required coupled P1 centers. The sample used (named Sample A in Table \ref{TableSample}) was a type Ib microdiamond powder synthesized by HPHT the method. According to its manufacturer, Element 6, this powder has a nominal P1 concentration between 110 and 130 ppm and a particle size between 15 and 25 $\mu$m. We verified the reproducibility of the DNP profile pattern with a different type of Ib and HPHT-made microdiamond powder (Sample B in Table \ref{TableSample}) of a different origin and different nominal P1 concentrations. 
Sample B had a lower nominal P1 concentration (between 10 and 100 ppm) and 
smaller particle size (6 to 12 $\mu$m) and was manufactured by Engis Corp. 
The same rich features for the $^{13}$C DNP frequency profile could be observed for both samples at 3.34 T, see SI Figure 1. 

\begin{table}[h]
\begin{tabular}{llll}
 Properties          & Sample A      & Sample B     \\\hline 
Diamond type         & Ib            & Ib           \\
Grown method         & HPHT          & HPHT         \\
Particles size       & $15-25\,\mu$m & $6-12\,\mu$m \\ 
P1 concentration     & $110-130$ ppm & $10-100$ ppm \\ 
Diamond manufacturer & Element 6     & Engis Corp   
\end{tabular}
\caption{Samples properties}
\label{TableSample}
\end{table}

We next repeated this experiment at 7 T using the same samples to examine the features of the $^{13}$C DNP frequency profile. Again, virtually indistinguishable DNP profiles are observed using both type Ib microdiamond powder samples (Table \ref{TableSample}), shown in Figure~\ref{DNP Samples}b. We see very complex and broad patterns, qualitatively similar to that at 3.34 T, that cannot be explained by $^{13}$C polarization from isolated electron spins alone. Furthermore, the lineshape and width of the DNP profile from isolated electron spins should reproduce the observed EPR manifold that has an FWHM of approximately 0.44 mT/12 MHz for the center line. Instead, we see a much broader pattern emerge that exceeds 50-100 MHz, suggesting that multiple effects contribute to the DNP profile, including contributions from multi-electron spin coupling.

The possible DNP mechanisms that contribute to the DNP profile are schematically illustrated in Figure \ref{DNP Samples}c concerning a hypothetical EPR line with an anisotropic broadening of the order of the nuclear Larmor frequency $\omega_n$. The most straightforward one-electron one-nucleus interaction would give rise to solid effect (SE) DNP which would show up as a positive and negative feature at $\pm\omega_n$ relative to the EPR feature. The cross effect (CE) and truncated cross effect (tCE)  DNP mechanisms require multi-electron spin coupling contributing to the polarization transfer. The cross effect~\cite{hwang_phenomenological_1967} requires two interacting electrons, whose resonance frequencies differ by $\omega_n$ and have unequal polarization (usually due to microwave irradiation at the frequency of one of the electron spins that leads to selective saturation) to transfer net polarization to the nuclear spins. The resultant DNP shape presents an antiphase profile, separated by $\omega_n$. The cross effect features typically fit within the EPR linewidth, unless there are additional EPR features that are separated from the center line by $\omega_n$. The truncated cross effect~\cite{equbal_truncated_2018} originating from the same electron spin source requires many (more than 3) coupled electrons interacting with a weakly coupled or isolated electron spin. The strongly coupled electron spins have to feature a broader and frequency-shifted EPR line (represented in green in Figure \ref{DNP Samples}c) relative to the spectrum of the weakly coupled electron spin (represented by the original EPR line in black). Due to fast relaxation of the strongly coupled electrons (in green), DNP only arises as a positive or negative peak, at the frequency of the narrow EPR line, depending on the relative position of the narrow EPR line with respect to the broadened EPR spectral density. In the case illustrated in Figure \ref{DNP Samples}c where the broadened EPR line is shifted towards lower frequencies relative to the original EPR spectrum, the tCE profile appears as a negative absorptive feature, while the presence of a broadened EPR line shifted towards a higher frequency would appear as a positive absorptive feature, see SI Figure 2. If the broadened EPR feature is strictly centered and symmetric with respect to the EPR spectrum of the isolated electron spin, which would be the case when the broadening is due to dipolar coupling that in a powder average gives rise to symmetric broadening, then the positive and negative tCE contributions would cancel out and yield no observable effect.

It is clear that the SE of the three lines does not add up to the observed DNP frequency profile. Even without spectral decomposition, it is apparent that there are DNP features broader than the EPR lineshapes of the P1 manifold. In order to understand what mechanisms are involved we need to decompose the DNP frequency profile into different mechanisms. Based on the prior success of decomposing the DNP frequency profile at 3.34 T into SE, CE, tCE, and apparent Overhauser Effect (OE) DNP mechanisms, we apply the same analysis to the DNP frequency profile acquired at 7 T.~\cite{shimon_large_2022}
Figure \ref{13C_DNP_DECOMPOSITION}a and b display the DNP spectrum at 3.34 T and 7 T respectively measured at room temperature (in black) and the resulting fit (in green). This fit has been obtained employing a rudimentary technique that involves convolving the experimental EPR line with functions that consist of one (for the OE and tCE) or two opposite signed delta functions separated by 2$\omega_n$ (for the SE and CE) to generate the fundamental profiles for the various DNP mechanisms. The amplitudes of these profiles were adjusted accordingly to attain an optimal agreement with the experimental spectrum. To construct the shapes, we individually processed each hyperfine manifold of the EPR line (except in the CE case, where all three manifolds were considered together), followed by fine-tuning the relative amplitudes. Please refer to section \ref{Decomposition} for a comprehensive explanation. The SE contribution to the DNP profile is shown in pink. As mentioned before, the resulting DNP spectrum should present an antiphase feature separated by twice the nuclear Larmor frequency (35 MHz and 74 MHz at 3.34 T and 7 T respectively) of $^{13}$C with their lineshape reflecting the EPR lineshape of each hyperfine manifold of the P1 center. The SE effect originates from isolated P1 defects (i.e., single electrons) that are interacting with $^{13}$C nuclei, or from weakly interacting P1 defects which do not fulfill the CE condition (i.e., their Larmor frequencies are not separated by the nuclear Larmor frequency)\cite{Hovav2010}. The CE contribution to the DNP profile is shown in red. The CE comes from the coupling of at least two electron spins. At 3.4 T the CE condition can be satisfied within a single nitrogen manifold, whereas at 7 T the CE condition requires the coupling between two hyperfine manifolds, i.e. one of the $\pm1$ manifold and the central line. Still, the contributions from SE and CE are not enough to explain the observed DNP features, so we have to also consider contributions from tCE. The tCE contribution to the DNP profile is shown in blue. As mentioned earlier, tCE requires an EPR spectral feature of P1 centers that are broad enough and shifted in frequency relative to the observed P1 center manifolds. Their observation requires the presence of broad clusters from coupled electron spins beyond two-electron spin interaction. A reasonable fit to the observed DNP profile, however, still requires the consideration of an additional apparent OE feature (shown in teal). The residual between the acquired spectrum and the fit markedly improves when the apparent OE is taken into account, see SI Figure 3. We will discuss the mechanistic origin of this effect later when demonstrating the cross-talk between P1s, but in short, the OE requires electron-nuclear Larmor fluctuations of the order of the electron Larmor frequency that is unlikely to be present in diamonds, at least as long as OE DNP is to rely on the traditional dynamic-mediated coupling between one electron and one nuclear spin. For now, we will refer to the negative DNP contribution from the central nitrogen manifold as an apparent OE. Here, we will first focus on the mechanistic origin of the DNP mechanisms that rely on multi-electron spin couplings. 
The decomposition of the DNP spectra clearly shows that there are strong contributions from CE and tCE that require not only coupled P1 centers but also the presence of strongly coupled and frequency-shifted clusters of P1 centers. The contributions from SE DNP are also strong, leading to our hypothesis that isolated P1 centers, and coupled and clustered P1 centers are all present in the type Ib diamond sample as distinct defect populations. Notably, the relative contribution of the SE at 7 T is lower than that at 3.4 T, as expected owing to its underlying forbidden transition. The dissimilarity in the relative contributions of cross-effect DNP between two fields can be readily understood by examining the disparity in the probability of electron-electron pairs at the $^{13}$C Larmor frequency apart. This is because the extent of contribution arising from g-anisotropy and hyperfine coupling in the EPR lineshape is different in different fields. 

Next, we design a series of experiments to establish the hypothesis that the isolated and clustered spins coexist as separate species and populations. Second, we seek direct spectroscopic evidence for the presence of clustered species. And third, we estimate the relative percentage of the clustered spins. To do so, we relied on the decomposition result (shown in Figure \ref{13C_DNP_DECOMPOSITION}a and b) that identified two specific frequencies at which the SE (magenta dashed lines) or the tCE (blue dashed lines) contributions are dominant with minimal contamination from other DNP mechanisms. The isolated mechanisms at play at those frequencies allow us to selectively explore the properties of the P1 species giving rise to those DNP mechanisms, such as the build-up time constants, see SI Figure 4. At both frequencies, distinct build-up rates were observed, as shown in Figure \ref{13C_DNP_DECOMPOSITION}c and d, with the SE presenting a much slower build-up trend compared to the tCE. We fit the build-up curve with a bi-exponential function, yielding two dominant time constants for each curve. The short build up time constant $T_{BU}^{\mathrm{short}}$ originates from the polarization of $^{13}$C nuclei spins coupled to an adjacent P1 center or multiple centers, while the long build up time constant $T_{BU}^{\mathrm{long}}$ is from the polarization build-up of more distant $^{13}$C spins that require a longer time due to nuclear spin diffusion. The time scale of the longest build up time depends on $^{13}$C $T_1$, which is on the order of several minutes to hours~\cite{Hoch1988,reynhardt_13c_1997}, and the slow nuclear spin diffusion constant in natural diamond (1.1\% $^{13}$C) equal to~\cite{reynhardt_13c_1997,terblanche_13c_2001} $D=0.67\times 10^{-15}\mathrm{cm}^2\,\mathrm{s}^{-1}$ (at 4.7 T and 300 K) to spatially distribute polarization originating from P1 centers across the sample.

At 3.34 T/7 T the tCE presents a $T_{BU}^{\mathrm{short}}$ of the order of 60s/30s and a $T_{BU}^{\mathrm{long}}$ of 7min/14min. The solid effect at 3.34 T/7 T presents a $T_{BU}^{\mathrm{short}}$ of the order of 1.5min/1min and a $T_{BU}^{\mathrm{long}}$ of 18min/35min. The disparity in the build-up times of the SE and tCE mechanisms corroborates that distinct electronic and/or nuclear spin environments are present and coexisting. However, the characterization of the NMR spectral linewidth and $T_2$ demonstrated no difference across the different EPR frequencies, as shown in SI Figure 5, implying that the $^{13}$C nuclear spins have identical properties and experience the same spin physical environment. {The results suggest that there are physically distinct electron spin environments contributing to the SE and tCE, while they contribute to the hyperpolarization of the entire $^{13}$C nuclear spin bath, as opposed to distinct populations of $^{13}$C  pockets.}
At 3.34 and 7 T, there is no EPR frequency at which the selective contributions from the CE can be examined without overlap from other DNP mechanisms, but we expect the buildup time of the P1 centers contributing to the CE to lie between that of the SE and tCE. 

These results confirm that clustered P1 centers represent physically distinct populations from the isolated P1 centers. This is surprising because in type Ib diamonds the possibility of P1-P1 interactions has been discussed in the literature~\cite{Samsonenko1965, loubser_electron_1978}, but the idea of P1 clusters impacting its DNP at a very high field (7T and above) is new. 

\subsection{EPR Experiments at 8.2 T}

\begin{figure}[h]
    \centering
    \includegraphics[width =0.8\textwidth]{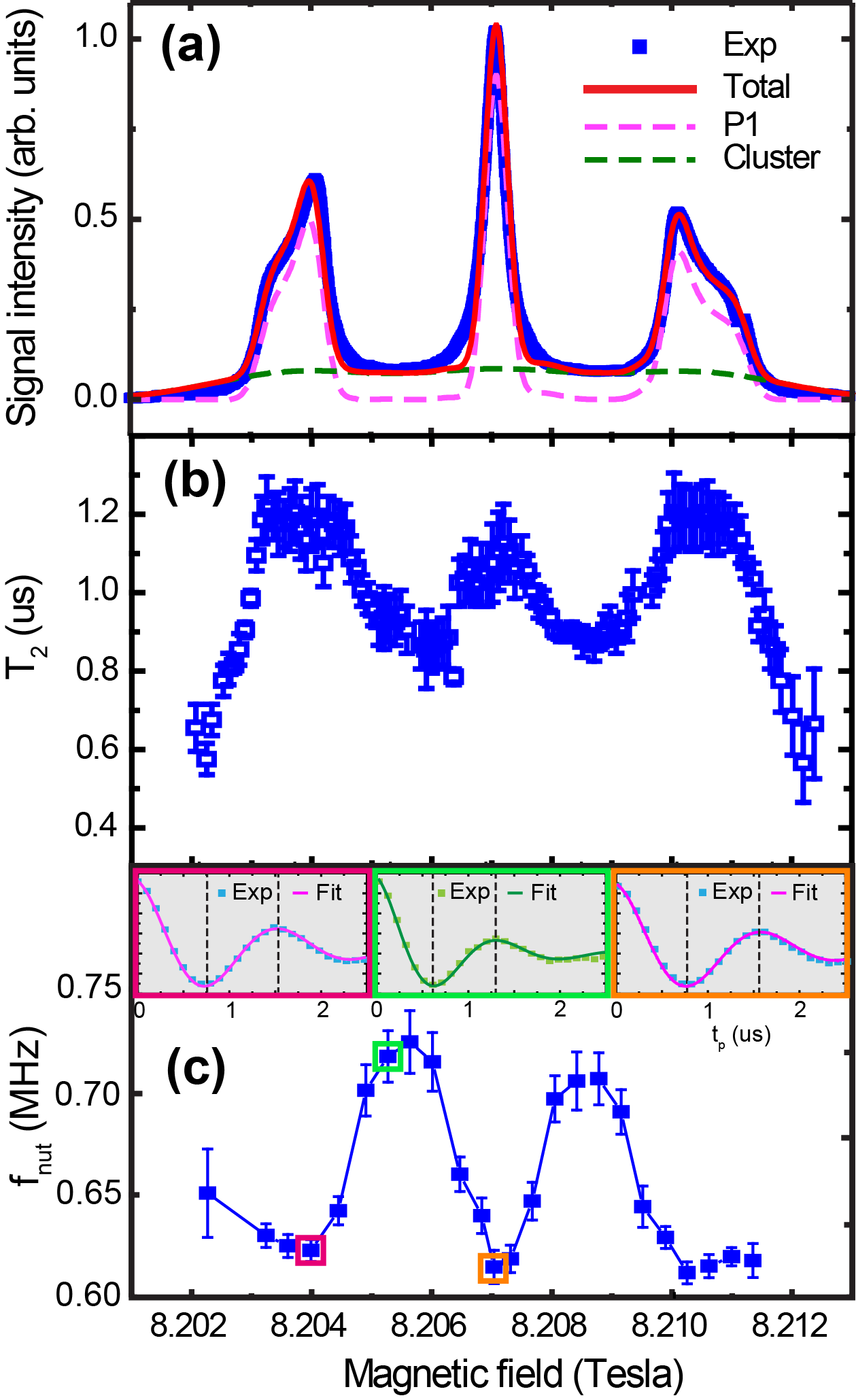}
    \caption{a) Echo-detected field-stepped EPR experiment at 8.2 T(blue squares) of sample A. EasySpin was used to fit (dashed line) using the P1 Hamiltonian with the following parameters: $S=1/2$, $g_x=g_y=2.0023$, $g_z=2.00225$, $A_x=A_y=82$ MHz, $A_z=114$ MHz and $Q=-4$ MHz and a combination of two different linewidths. For the isolated P1 centers (purple dashed line) the linewidth was set to 0.33 mT, while for the clustered specie (green dashed line) it was set to 2.7 mT. b) Transverse electronic relaxation time ($T_{2e}$) as a function of magnetic fields. A monoexponential decay was fitted to extract the $T_{2e}$. The error bars represent 95 \% confidence intervals.  
    c) Nutation frequency as a function of magnetic fields. A nutation measurement sequence ($t_p - T - t_{\pi/2} - \tau - t_{\pi} - \tau - echo$) was used. $T=10$ $\mu$s, $t_{pi/2} = 300$ ns, $t_{\pi} = 500$ ns and $\tau=850$ ns. In the measurement, the echo intensity was measured while the pulse duration of $t_p$ was varied as shown in the inset. The inset shows the measurement data taken at 8.2041, 8.2054 and 8.2071 T. The nutation frequencies ($f_{nut}$) were extracted by fitting the data with $A \cos(2\pi f_{nut} t_p) + y_0$ where $f_{nut}$, $A$ and $y_0$ are fitting parameters. The EPR spectra of sample B are shown in the SI Figure 7}
    \label{EPR_Susumu}
\end{figure}

{ As discussed earlier, tCE DNP requires the presence of clusters whose EPR line is broad and shifted in its center frequency.} If so, we should be able to directly detect such populations by performing EPR experiments with short echo delays and/or at cryogenic temperatures. In an attempt to observe clustered P1 centers, we conducted an echo-detected field-stepped EPR experiment at room temperature and 8.2 T using an echo delay of $2\tau=1.7$ $\mu$s. Figure \ref{EPR_Susumu}a displays the EPR line of the P1 center sample. We see the expected hyperfine manifold of the P1 center but with a clearly elevated baseline (the same characteristics are observed with Sample A shown here and sample B shown in SI Figure 7). To verify the existence of two spectral components, we employed EasySpin~\cite{STOLL200642} to fit the spectrum with two components. We obtain a fit with high confidence with two populations. 
We utilized the same Hamiltonian for both environments but with different widths, taking into account the dipolar broadening of clustered spins.
The following is the employed Hamiltonian in unit of $s^{-1}$.
\begin{equation}
\hat{H} = \frac{\mu_B}{h} \hat{S} \cdot \stackrel{\leftrightarrow}{g} \cdot \vec{B_0} + \hat{S} \cdot \stackrel{\leftrightarrow}{A} \cdot{I},
\end{equation}
where $\mu_B$ is the Bohr magneton, $h$ is the Planck constant, $\hat{S}$ and $\hat{I}$ are the spin operators for P1 electron and $^{14}$N nuclear spins, respectively. $B_0$ is the magnetic field. $\stackrel{\leftrightarrow}{g}$
is the $g$-tensor. $\stackrel{\leftrightarrow}{A}$
is the hyperfine tensor.

The experimental parameters used in our fitting were first established by simulating the narrow EPR manifold that yielded the following values: spin $S=1/2$, $g$-tensor values $g_x=g_y=2.0023$, and $g_z=2.00225$, hyperfine couplings $A_x=A_y=82$ MHz, $A_z=114$ MHz, and quadrupole coupling constant $Q=-4$ MHz.
This $g$-anisotropy alone gives rise to a full-width half maximum of 0.44 mT (corresponding to 12 MHz) of the central EPR line and is responsible for the asymmetric shape and the uneven amplitudes of the $m_s=\pm$1 manifolds.
Given this, the linewidth of the isolated P1 centers was set at 0.44 mT, taking $g$-anisotropy into account. The broad EPR component that we attribute to clustered P1 centers was fit and yielded a linewidth of 2.7 mT, which corresponds to a line broadening of 76 MHz. Both, the 12 MHz linewidth due to $g$-anisotropy and the 76 MHz broadening of the clustered P1 centers are significantly larger than that obtained from Monte Carlo simulations of distributions of dipolar couplings between P1 centers that are statistically distributed in the diamond lattice at concentrations ranging between 20 ppm and 160 ppm (see SI Figure 6). At 20 ppm, 83\%  of the couplings are between 0.75 MHz, while at 160 ppm, 80\%  of the couplings are between 1.95 MHz, i.e. the dipolar broadening due to statistical distribution of P1 centers is miniscule. If we take 76 MHz broadening of the narrow EPR line as the isotropic dipolar coupling value ($(\mu_0/(4\pi)(g^2\mu_B^2/h) m_z^2/d^3$ where $m_z=\pm 1/2$) for the sake of argument, that value would originate from a P1-P1 distance of $d=0.56$ nm. { While this value is dramatically shorter compared to the P1-P1 distance for isotropic distribution at the nominal P1 concentration,} this value is still much greater than the lattice constant of diamond of $a=0.36$, suggesting that the coupling seen here that gives rise to the distinct DNP properties do not necessarily need to originate from adjacent defects or other color centers. The amplitudes of the peaks were adjusted to obtain the best fit for the experimental data and show that the contribution of P1 clusters is at least 30\% of that of the isolated P1 centers (or 23\% of the total P1). Note that broader EPR lines would not be readily measurable, making this estimate a lower limit for the P1 cluster population. Also, the microsecond-scale delays between the pulses are expected to diminish the fast-relaxing spins due to strong spin interactions. 
Of course, line broadening is an effect averaged over a distribution that includes contributions from stronger and weaker couplings. Critically, when exchange coupling is involved, the relationship between broadening and distance is no longer straightforward. Hence, further clarifications on the nature of the P1 clusters need to rely on measurement methods beyond EPR lineshape analysis.

As mentioned earlier, clustered spins should exhibit a fast relaxation decay, which prompted us to investigate the electronic transverse relaxation time ($T_{2e}$) as a function of field. We measured the spin echo decay across the entire P1 manifold, {\it i.e.} at different magnetic fields, and fitted a monoexponential function to extract $T_{2e}$. The resulting $T_{2e}$ values are shown in Figure \ref{EPR_Susumu}b. This figure not only clearly illustrates the presence of two different $T_{2e}$ values, but more importantly, that the EPR spectral density in between the hyperfine manifolds that would be typically attributed to a baseline shows consistently much shorter $T_{2e}$ values compared to that of the three main hyperfine EPR signals. The two coexisting distinct spin environments feature longer $T_{2e}^{\mathrm{P1}}\approx1.07\,\mu$s for the isolated P1 centers and shorter $T_{2e}^{\mathrm{Cluster}}\approx0.88\,\mu$s for the clustered P1 centers. So, the EPR spectral density in between the main hyperfine manifold is not a baseline but originates from clustered P1 populations. 
{ Similar to Sample A, the contribution of clustered spins was also observed in the EPR spectra and $T_2$ of Sample B (Supporting Information).}
We furthermore estimate the spin concentration of P1 centers using $T_2$ of P1 centers and Eq.(11) in the previous work~\cite{stepanov_determination_2016}.
For Sample A, with $T_2 = 1.07 \pm 0.09$ $\mu$s, we determined the P1 concentration to correspond to $67 \pm 7$ ppm. 
We note that this value excludes the concentration of the spin clusters, and hence are reporting on the non-clustered P1 centers. As shown in Table~\ref{TableSample}, the so estimated obtained P1 concentration is much lower than the nominal P1 concentration provided by the manufacturer, Element 6. This result is consistent with the existence of a considerable amount of spin clusters in Sample A that are not included in the $T_2$-based concentration calibration.
{Similarly, we determined that the P1 spin concentration of Sample B is $47 \pm 3$ ppm using the previously measured $T_2$ value of Sample B~\cite{Peng_2020} and Ref.~\cite{stepanov_determination_2016} (shown in SI Figure 7). 
As shown in Table~\ref{TableSample}, the result is still within the range of the manufacturer's specification.}

Next, we performed nutation experiments as a function of frequency across the entire EPR manifold. The objective was to investigate whether the clustered P1 centers have high-spin character and/or experience significant exchange coupling, besides dipolar coupling. To our surprise, the nutation frequency was significantly elevated for the P1 population in between the hyperfine manifolds (Figure \ref{EPR_Susumu}c), suggesting that a significant population of P1 centers displays spin quantum numbers of 1 or greater due to clustering. Notably, when two strongly exchange coupled spins are present, their behavior resembles that of a spin 1 system. As a result, the nutation of these coupled spins occurs at a faster rate. For example, the nutation frequency of a pair of strongly exchange coupled spin 1/2 spins will be greater by a factor of ($\sqrt{2}$) than that of an isolated spin 1/2 system. As shown in Figure \ref{EPR_Susumu}c, we indeed observe a significantly faster nutation frequency for the clustered population by a factor of 1.25 compared to that of the non-clustered P1 centers. The clustered P1 centers are likely a mixture of coupled spins that have spin 1/2 and spin 1 or greater quantum numbers

In contrast, there is no contribution from isolated P1 centers between the peaks, making it easier to establish the different contributions. This EPR analysis establishes that this "invisible" EPR line from clustered species is a significant contribution that makes up at least 30\% with respect to the isolated P1 clusters. Still, the resolution of echo-detected EPR measurement is insufficient to identify the exact spectral features underlying the broad baseline. The broad feature may consist of multiple contributions, contain asymmetric spectral features, and very likely makes up a larger population than estimated with the tools available for this study. { We also note that the broad baseline features in the echo-detected EPR spectrum cannot be $a priori$ assigned to P1 clusters without prior knowledge, given that many paramagnetic impurities can give rise to such broadening. In this study, DNP characterization established that clustered P1 engage in cross-talk with isolated P1 centers and make up a non-negligible population of the paramagnetic centers, given the strong CE and tCE signatures.}

\subsection{Pump-probe EPR Experiments}

\begin{figure*}[h]
    \centering
    \includegraphics[width =\textwidth]{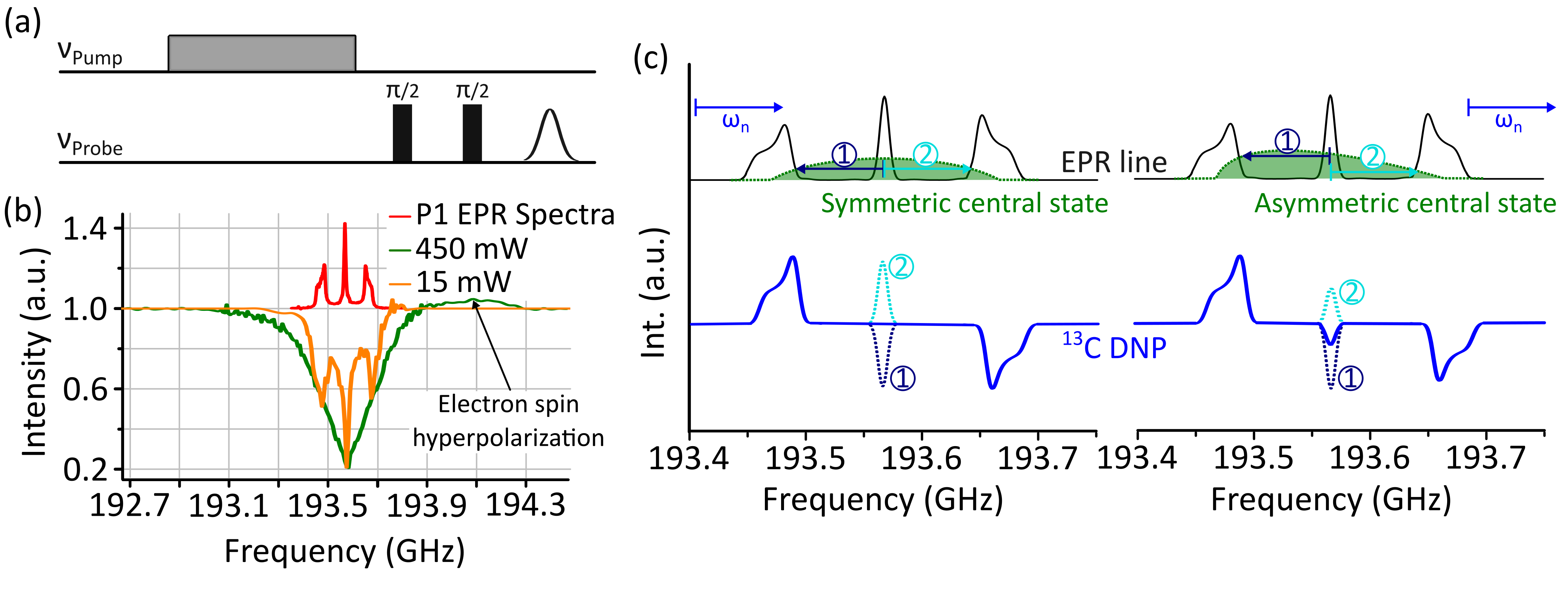}
    \caption{a) Electron-electron double resonance (ELDOR) pulse sequence. A long pump mW pulse (in the order of the ms) with variable frequency ($\nu_{\mathrm{pump}}$) is applied after detecting with a solid echo at a defined probe frequency ($\nu_{\mathrm{probe}}$). b) ELDOR spectra with two different pump microwave power detecting at $\nu_{\mathrm{probe}}=$ 193.566 GHz. Low power (15 mW) is shown in orange while high power (450 mW) is shown in green. The high-power ELDOR presents a much broader and asymmetric dip compared with the same at low power. As a reference, the P1 EPR line is shown in red. An arrow indicates the electron spin hyperpolarization on the high power ELDOR profile. c), Schematic of a symmetric (left) and asymmetric (right) central broad feature (highlighted in green) with respect to the central EPR peak (shown in black). The interaction with the nuclear Larmor frequency $\omega_n$ of the central EPR peak with the broad EPR feature to the left/right generates a negative/positive DNP peak (defined as 1/2, shown in blue/light blue dashed lines). If the central broad line is symmetric/asymmetric there is no/negative net DNP central polarization and the side DNP peaks have equal/different intensities (left/right schematic). }
    \label{ELDOR}
\end{figure*}

As mentioned earlier, for tCE to occur the P1 centers must exhibit a broad and frequency-shifted EPR spectral characteristic. Additionally, if the broadened EPR feature is precisely centered and symmetrical compared to the isolated electron spin's EPR spectrum, the positive and negative tCE contributions would cancel each other, resulting in no detectable effect. In order to understand whether the broad EPR features have multiple, frequency shifted and/or asymmetric features with respect to the central P1 center EPR line, we can use a multi-frequency pump-probe EPR technique that measures the extent of saturation and cross talk between the electron spin populations. For this, we utilize electron-electron double resonance (ELDOR), a method to detect the response of the EPR echo intensity at a given probe frequency ($\nu_\mathrm{probe}$) upon applying a pump pulse prior (pulse sequence shown in Figure \ref{ELDOR}a). This experiment was carried out at 7 T and 80 K in order to slow down relaxation compared to room temperature to more readily observe the potential cross-talk between the different P1 center populations. We applied a series of pump pulses at different EPR frequencies ($\nu_\mathrm{pump}$) along the P1 spectrum and detected at the probe frequency corresponding to the center frequency of $\nu_\mathrm{probe}=$ 193.566 GHz, shown in Figure \ref{ELDOR}b. At a low pump power of 15 mW, we see that the ELDOR spectrum is inverted (shown in orange) with respect to the EPR spectrum (in red), which indicates that the saturation profile of the EPR spectrum roughly reflects the shape of the EPR spectrum with modest line broadening. When applying higher microwave power of 450 mW (corresponding to the power applied to observe $^{13}$C DNP), we observe a dramatically broadened ELDOR profile with the hyperfine features entirely obscured, shown in green in Figure \ref{ELDOR}b.
Furthermore, the broadening of the ELDOR profile is asymmetric (the ELDOR profile shows greater saturation at lower pump frequency), indicating that there are electron spin populations that give rise to a broad and frequency-shifted spectral density (exactly the conditions required to give rise to tCE). Even more surprisingly, we observe the asymmetric ELDOR profile with an elevated echo intensity at the higher frequency side. Such electron spin hyperpolarization (the detected echo has a higher amplitude than the echo measured without applying a pump pulse) has been previously shown to be a characteristic signature of clustered electron spin populations by Equbal et al.~\cite{equbal_truncated_2018}, and predicted in studies of Vega and coworkers~\cite{kundu_theoretical_2019} and Atsarkin and coworkers~\cite{atsarkin_spin_2017,atsarkin_temperature_1972,atsarkin_dynamic_2012}. In conclusion, we provide ample evidence for the existence of clustered, exchange coupled, P1 centers that not only give rise to broad spectral features with shorter $T_{2e}$ relaxation times but also electron spin resonance features that are frequency shifted. The origin of a frequency shift can be a strong dipolar coupling of P1 centers with g anisotropy and/or the presence of strong exchange coupling exceeding 80 MHz (corresponding to the $^{13}$C nuclear Larmor frequency) with a positive or negative sign. However, the precise analysis of such features is not trivial and would require DNP and ELDOR measurements at even higher magnetic fields than 3.34 and 7 Tesla, as provided by concurrent studies performed in the lab of Kaminker (private communication). 

To complete our analysis, we now need to come back to the decomposition of the DNP profile that required the consideration of the apparent OE in previous studies, i.e. fluctuations and dynamics. Because we did not think the OE originating from one electron - one nuclear spin coupling (the common spin system assumed for OE) is plausible in diamond samples, we sought an alternative explanation. 
 The physical origin of tCE and OE is entirely different. As mentioned, the tCE requires the cross-talk between isolated and clustered P1 centers, while OE typically relies on cross-relaxation between one electron spin coupled to one nuclear spin induced by spin density fluctuations of either the P1 center or the $^{13}$C nucleus. Although their physical implications are different, the tCE can mimic features of an OE due to the appearance of an isolated peak in a particular energy crossing frequency. Hence, a priori differentiating between them is difficult by the shape of the DNP profile alone. 
We obtained hints that the apparent OE might also be originating from clustered electron spins based on the DNP build up time constant at the center frequency of 193.566 GHz at which the apparent OE observed build up time approaches that of the tCE. Our hypothesis is that the apparent OE in the central part of the DNP profile at both fields, 3.34 T and 7 T, is in fact a tCE between the slow-relaxing central EPR peak of the P1 center and a fast-relaxing and frequency-shifted spectral feature. The existence of a broad and asymmetric spectral component was confirmed in the previous paragraph, while we know that there is cross-talk between the two outer P1 manifolds and the slow relaxing EPR features that give rise to the truncated cross effect. It is not logical that tCE features are seen at the positions of the outer EPR manifolds, but not at the frequency of the central EPR peak. We present a plausible mechanism for this observation.  For that, we will use the schematic drawn at the top of Figure \ref{ELDOR}c. The EPR line of the isolated P1 center is shown in black, along with (an exaggerated) broad central EPR feature, shown in green, illustrated as a symmetric shape (left) or an asymmetric shape (right) with respect to the central EPR peak. The span of the nuclear Larmor frequency ($\omega_n$) is highlighted with a blue arrow. The resulting $^{13}$C DNP profile, considering only the tCE effect, is shown in Figure \ref{ELDOR}c, bottom. The interaction of the central EPR peak with the symmetric broad EPR feature will yield a negative and positive DNP profile of equal intensities (1 and 2 respectively on the schematic) so they will cancel out leaving no net $^{13}$C polarization. 
In order for a net negative DNP peak intensity to arise, as observed at 3.34 T and 7 T, the central broad EPR feature has to present an asymmetry with the higher EPR intensity on the lower frequency side. This will result in a slightly more intense negative peak than the positive peak on the $^{13}$C DNP profile. Furthermore, if this is true, the intensities of the tCE DNP features of the outer manifolds should be different. This is precisely what we observe on the DNP profile at both magnetic fields (see Figure \ref{13C_DNP_DECOMPOSITION}a and b). 
The presence of a broad asymmetric dip of the ELDOR spectrum, along with the agreement of the $^{13}$C DNP profile with our hypothesized schematic present sufficient proof for the assignment of the apparent OE as the tCE. With this, all observed DNP features can be rationally assigned to a mechanism. The dominant contribution of the tCE implies that there is significant exchange coupling between the clustered P1 centers that give rise to a frequency shift of a broad spectral feature, corroborated by an enhanced nutation frequency of this population, and cross talk between the clustered and non-clustered P1 centers. This finding also demonstrates that a characteristic, OE DNP-like, absorptive component in the DNP profile can serve as a signature for strongly clustered P1 centers. 

\section{Conclusions}

Using $^{13}$C  DNP of P1 centers, we were able to identify the existence of a strongly clustered P1 population that makes up at least 30\% of isolated P1 centers in type Ib diamonds. Remarkably, this property of a strongly heterogeneous P1 center distribution is independent of the manufacturer and nominal P1 concentrations in type Ib diamonds synthesized by HPHT. Whether or not strong clustering of P1 centers occurs with CVD diamonds remains to be seen. However, we speculate that such clusters exist also in CVD diamonds, albeit at lower concentrations, because the clustering must be a result of intrinsic property of substitutional nitrogen defect formation, migration, and annealing energetics.

The unexpected discovery of such a fundamental property of diamonds is surprising, considering the large community studying these materials for quantum information applications for which the spatial distribution and spin physics properties of P1 and NV centers are paramount. We believe that this critical property remained unknown and/or did not receive much attention from the diamond quantum sensing community because the EPR features of clustered P1 centers (and for that matter NV centers) are hard to see at low or earth magnetic fields where the majority of studies are performed. The presence of inhomogeneously distributed P1 clusters will give rise to a broad range of experimentally-measured T$_2$ times in single site NV measurements \cite{li_determination_2021} as well as spatially-dependent results for studies of the local magnetic field fluctuations detected via noise spectroscopy \cite{romach_2019}, but their direct detection has been elusive and their existence has not received much attention in the literature.

{In contrast, $^{13}$C DNP profile measurements revealed strong features attributed to clustered P1 centers that could not be ignored.}
It turns out that $^{13}$C DNP at high magnetic field is an exceptionally sensitive tool to reveal the existence of P1 (or any paramagnetic) clusters with electron-electron (D or J) coupling larger than the $^{13}$C nuclear Larmor frequency, which is 35 MHz at 3.34 Tesla and 74 MHz at 7 Tesla. These frequencies correspond to isotropic electron dipolar coupling of the order of 1 nm distances (physical attribution to J coupling is more challenging) but are less apparent in an EPR spectrum acquired at low magnetic field and difficult to see in a CW EPR spectrum due to saturation effects. When equipped with the knowledge that a sizable population of P1 centers is clustered, pulsed echo-detected EPR experiments can be used to identify the signatures of strongly coupled P1 clusters that have broad EPR features, shorter T$_{2e}$ relaxation time, and also greater nutation frequencies. 
The observation of greater nutation frequencies for the clustered P1 centers proves that there exists strong e-e exchange coupling between the P1 centers that, remarkably, give rise to high-spin properties.
 We furthermore confirmed by $^{13}$C DNP profile and ELDOR measurements that the broad EPR species that we attribute to clustered P1 centers not only are prevalent, but also engage in strong cross-talks with the isolated P1 centers, making them highly relevant for the spin dynamic properties of the entire P1 and NV center population.  { While these experiments benefit from high magnetic field, they can also be performed at lower field, once the knowledge is established where to look and what to look for.}

We propose that room temperature $^{13}$C DNP profile measurements at high magnetic fields should serve as a valuable tool for characterizing of P1 defects as well as NV centers in various diamond samples. The construction and operation of such a DNP instrument can be readily achieved with modifications to an existing solution-state NMR system. Furthermore, no cryostat operation is needed as all the DNP experiments were performed at room temperature and a solid-state microwave source can be used~\cite{tagami_design_2023,guy_design_2015,feintuch_dynamic_2011}. This DNP profiling technique offers a non-invasive and reliable means of assessing the electron spin populations and electron spin clustered, aiding in the understanding of diamond defects and their potential applications in quantum technologies. This does not mean that the applications of quantum technologies should occur at high magnetic field. Rather, electron and nuclear spin resonance spectroscopies at high magnetic field have the potential to fuel fundamental discoveries of materials and spin physics properties of spin-based quantum sensors and quantum qubit candidates that have been obscured or overlooked and can guide us toward new or optimal conditions to perform quantum sensing or quantum computing. { Last, but not least, the discovery of high population of exchange coupled P1 spin clusters is highly relevant for DNP-enhanced NMR or MRI applications of intrinsic $^{13}$C NMR signal or NMR sensing applications of external analytes, given the high DNP enhancements achieved at room temperature using low microwave powers from a solid-state source. There is significant room through diamond material processing geared towards optimizing the targeted DNP mechanisms, while this study offers guidelines what spin properties one should look out for.}

\section{Methods}
\subsection{Samples}
For all measurements presented in this study, we used a powder diamond sample donated by Element 6. The type Ib diamond is made by HPHT synthesis. The diamond microparticles are 15-25$\mu$m in diameter and are specified to have a nitrogen concentration of 110 to 130 ppm. To compare the $^{13}$C DNP profile we also utilized a type Ib, HPHT, microdiamond powder manufactured by Engis Corp. with microparticle size between 6 to 12$\mu$m and concentration between 10 to 100 ppm. 

\subsection{7 T DNP, EPR, and ELDOR Measurements}
High-field EPR and DNP measurements at 7 T were performed at the University of California Santa Barbara. The 7 T dual EPR/DNP system is a home-built and cryogen-free system as described in more detail in a previous publication~\cite{leavesley_versatile_2018}. The system includes a Bruker superconducting magnet with a 300 MHz, 89 mm wide bore, a Bruker Avance D300WB console, and a 1 kW radio frequency amplifier for the X-nuclei channel. We conducted the experiments using a solid-state MW source and a 200 GHz MW transmitter system, which consisted of a 12 GHz Yttrium Iron Garnet (YIG)-based synthesizer and a 16x amplifier-multiplier chain (AMC), powered by an amplifier (VDI AMC629). This system is tunable over an 8 GHz frequency range of 182-200 GHz and has a microwave power output of up to 450 mW. We control the microwave frequency of the 12 GHz YIG-synthesizer using a PC and the Specman4EPR software, which allowed us to achieve a 1 Hz resolution and a response time of fewer than 12 ms. The MW output was transmitted through a 12.5 mm ID transmission horn and controlled by a quasi-optical bridge before being directed downwards by a corrugated waveguide, which was built into the probe that was placed within the 7 T magnet. The sample was held at the bottom of the waveguide by a "J-arm" structure. We used a custom-designed inductively coupled double resonance NMR circuit ($^1$H - $^{13}$C) in the probe to achieve high NMR/EPR performance at cryogenic temperatures, following a design principle developed by Tagami and Zens et al.~\cite{zens_using_2020}.

Echo-detected frequency sweep experiments were performed using two 700 ns pulses separated by 400 ns to detect a solid echo. $^{13}$C DNP frequency profile experiments used 3000 s of microwave irradiation duration at maximum power (450 mW). 81 frequencies were used between 193.37 GHz and 193.77 GHz. An FID was detected after microwave irradiation using a pulse of 35 $\mu$s. $^{13}$C DNP build-up experiments were measured using the maximum available microwave power.

For ELDOR experiments, we used a second microwave synthesizer (VDI Inc. VDIS 0060) with frequency sweep and frequency modulation capability. The microwave power was amplified to around 60 mW using an AMC (VDI Tx233) in a frequency range tunable between 190 GHz and 200 GHz. We used a pair of fast ($\sim$100 MHz) TTL-controlled pin switches in the VDI transmitter system to select between the output of the two sources. To carry out low-temperature ELDOR experiments, we used a custom-designed closed-cycle cryostat that was located inside the bore of a magnet. The cryostat had a hollow bore surrounded by a vacuum jacket, and the DNP probe was loaded and secured with a KF-50 clamp to create a vacuum seal. We used a Sumitomo CKW-21 helium compressor and a Janis custom SHI-500t-5 Gifford-McMahon cold head to cool the sample by gas conduction. The sample chamber was filled with helium gas, and the cryostat was pressurized at 0.5 psi using an external helium tank, a pressure sensor, and a solenoid valve. 
ELDOR experiments used a 35 ms irradiation pulse with two different microwave power settings (450 mW or 15 mW) and then detected with the second mw power source using two 850 ns pulses separated by 400 ns to detect a solid echo. 

\subsection{8.2 T EPR Measurements}
Pulsed EPR experiment was performed at 8.2 Tesla using a 230 GHz EPR spectrometer built at the University of Southern California. The EPR spectrometer consists of a high-frequency high-power solid-state source, quasi optics, a corrugated waveguide, a 12.1 T EPR sweepable superconducting magnet, and a superheterodyne detection system. The output power of the source system is 100 mW at 230 GHz. A sample on a metallic end-plate at the end of the corrugated waveguide is placed at the center of the superconducting magnet. Details of the system have been described elsewhere ~\cite{cho_high-frequency_2014}. In the present study, the diamond powder sample was placed in a Teflon sample holder (5 mm diameter), typically containing 5 mg of diamond powder.~\cite{cho_230115_2015,peng_investigation_2019}

\subsection{3.4 T DNP Measurements}
The 3.4 T DNP experiments, including the DNP spectrum and buildup curves, as well as the fitting procedure were described in detail in a recent publication by Shimon et al.~\cite{shimon_large_2022}

\subsection{Fitting the 7 T DNP Frequency Profile}
\label{Decomposition}
The fitting procedure described by Shimon et al.~\cite{shimon_large_2022} and used at 3.4 T was revised here and adapted for use at 7 T.

Briefly, we fit the DNP spectrum with a linear combination of crudely simulated DNP spectra for each one of the DNP mechanisms, changing their amplitude in order to get a good fit for the experimental DNP spectrum.

First, the EPR line used for fitting was simulated at 7 T using a single e-$^{14}$N spin system, as described in the main text, using EasySpin~\cite{stoll_easyspin_2006}. For the Solid Effect, Apparent Overhauser Effect, and Truncated Cross Effect, the EPR line was split into three according to the nitrogen spin manifold. This allowed for an independent contribution of each mechanism from each nitrogen spin manifold. Because the appearance of CE is expected between the three nitrogen spin manifolds, the EPR line was not separated into three parts when calculating the CE-DNP lineshape but used as a single line that includes all three nitrogen spin manifolds.

Next, the shapes for the different DNP mechanisms were then calculated by convoluting a single delta function or two opposite signed delta functions separated by $\omega_n$ for each mechanism with the EPR line, exactly as described previously~\cite{shimon_large_2022}. A Gaussian broadening of 0.15 MHz was applied to the Solid Effect, and 12 MHz to the apparent Overhauser Effect, Cross Effect, and Truncated Cross Effect shapes. The larger broadening compared to the values used at 3.4 T was used to mimic the larger microwave power available with the 7 T DNP spectrometer. Finally, the broadened DNP shapes were used to fit the experimental data. The best fit was decided by eye.

\section*{Supporting Information}
$^{13}$C DNP profiles of two type Ib microdiamonds at 3.34 T and room temperature,	truncated cross effect appearance depends on the symmetry of the broad line, residual of the DNP fit with and without OE,	$^{13}$C DNP build up time at different microwave irradiation frequencies, $^{13}$C NMR T$_2$ and linewidth show that the $^{13}$C ensemble is uniform, histograms of dipolar interactions between statistically distributed P1 centers

\section*{Acknowledgement}
Contributions from SB and SH were supported by the National Science Foundation (NSF) grant \#2004217.
Contributions from YR and ST were supported by the NSF (ECCS-2204667 and CHE-2004252 with partial co-funding from the Quantum Information Science program in the Division of Physics), the USC Anton B. Burg Foundation, and the Searle scholars program (ST). The contribution from CR was supported by the NSF under cooperative agreement \#1921199 and grant \#2203681. Contribution from AE was supported by Tamkeen under the NYU Abu Dhabi Research Institute grant \#CG008.
We thank Ilia Kaminker for fruitful discussions. 

\begin{figure}[h]
    \centering
    \includegraphics[width =0.95\textwidth]{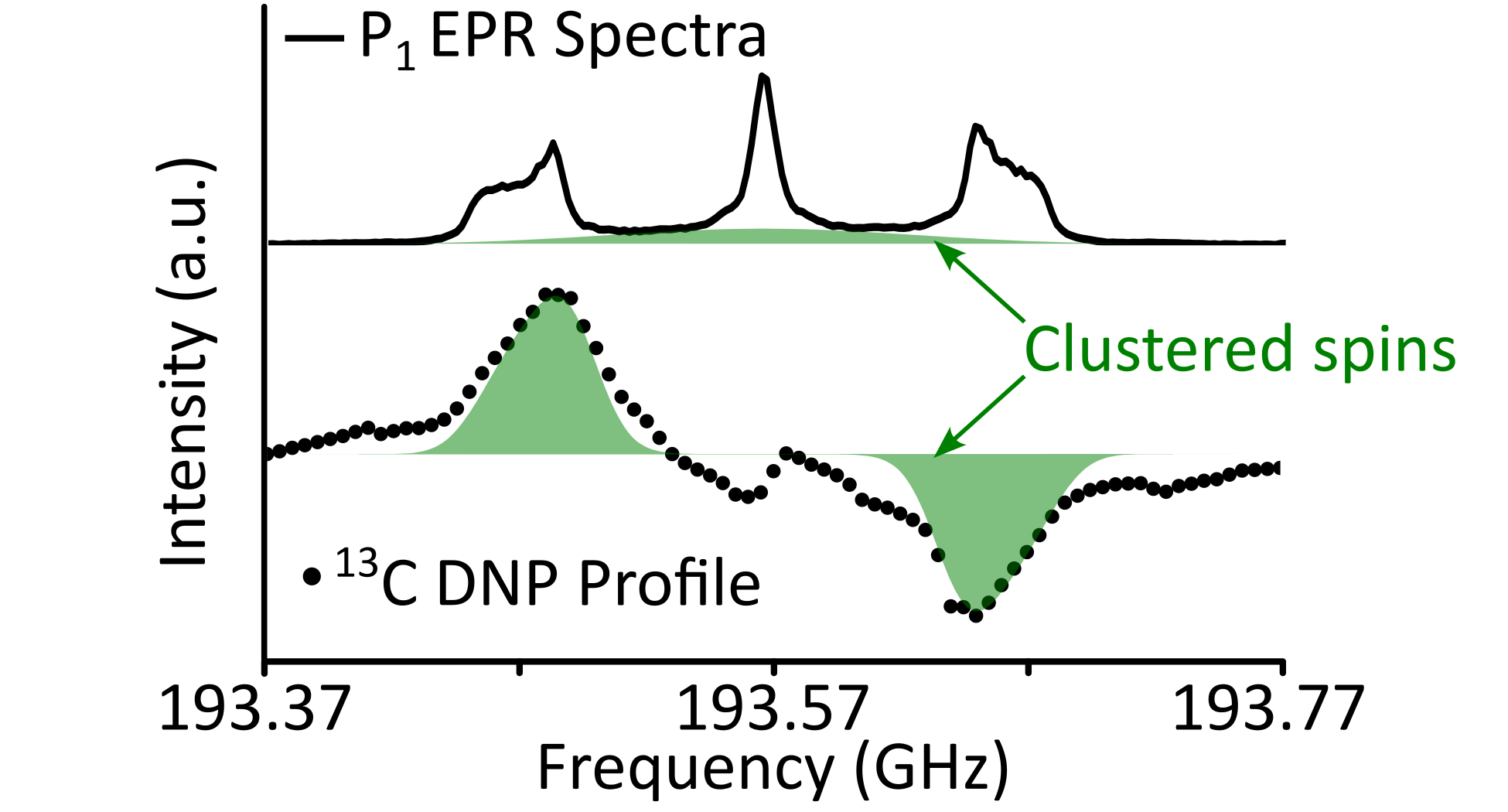}
    \caption{TOC graphic}
\end{figure}

 \bibliographystyle{files_latex/elsarticle-num-names-no-url}

\bibliography{ReferenceP1.bib}
\end{document}